\documentclass[12pt]{article}
\usepackage{html}
\usepackage{epsf}
\begin{document}
\title{MATTERS OF GRAVITY, The newsletter of the APS Topical Group on 
Gravitation}
\begin{center}
{ \Large {\bf MATTERS OF GRAVITY}}\\ 
\bigskip
\hrule
\medskip
{The newsletter of the Topical Group on Gravitation of the American Physical 
Society}\\
\medskip
{\bf Number 28 \hfill Fall 2006}
\end{center}
\begin{flushleft}
\tableofcontents
\vfill
\section*{\noindent  Editor\hfill}
David Garfinkle\\
\smallskip
Department of Physics
Oakland University
Rochester, MI 48309\\
Phone: (248) 370-3411\\
Internet: 
\htmladdnormallink{\protect {\tt{garfinkl-at-oakland.edu}}}
{mailto:garfinkl@oakland.edu}\\
WWW: \htmladdnormallink
{\protect {\tt{http://www.oakland.edu/physics/physics\textunderscore people/faculty/Garfinkle.htm}}}
{http://www.oakland.edu/physics/physics_people/faculty/Garfinkle.htm}\\

\section*{\noindent  Associate Editor\hfill}
Greg Comer\\
\smallskip
Department of Physics and Center for Fluids at All Scales,\\
St. Louis University,
St. Louis, MO 63103\\
Phone: (314) 977-8432\\
Internet:
\htmladdnormallink{\protect {\tt{comergl-at-slu.edu}}}
{mailto:comergl@slu.edu}\\
WWW: \htmladdnormallink{\protect {\tt{http://www.slu.edu/colleges/AS/physics/profs/comer.html}}}
{http://www.slu.edu//colleges/AS/physics/profs/comer.html}\\
\bigskip
\hfill ISSN: 1527-3431

\begin{rawhtml}
<P>
<BR><HR><P>
\end{rawhtml}
\end{flushleft}
\pagebreak
\section*{Editorial}
It is hard to imagine Matters of Gravity without Jorge Pullin as 
editor, and yet here it is.  Greg Comer and I are new at this and 
we could use your help.  In particular, if you have ideas for topics
that should be covered by the newsletter, please email us and/or 
the relevant correspondent.  Any comments/questions/complaints
about the newsletter should be emailed to me.

The next newsletter is due February 1st.  This and all subsequent
issues will be available on the web at
\htmladdnormallink 
{\protect {\tt {http://www.oakland.edu/physics/Gravity.htm}}}
{http://www.oakland.edu/physics/Gravity.htm} 
All previous issues are available at
\htmladdnormallink {\protect {\tt {http://www.phys.lsu.edu/mog}}}
{http://www.phys.lsu.edu/mog}
A hardcopy of the newsletter is distributed free of charge to the
members of the APS Topical Group on Gravitation upon request (the
default distribution form is via the web) to the secretary of the
Topical Group.  It is considered a lack of etiquette to ask me to mail
you hard copies of the newsletter unless you have exhausted all your
resources to get your copy otherwise.

\hfill David Garfinkle 

\bigbreak

\vspace{-0.8cm}
\parskip=0pt
\section*{Correspondents of Matters of Gravity}
\begin{itemize}
\setlength{\itemsep}{-5pt}
\setlength{\parsep}{0pt}
\item John Friedman and Kip Thorne: Relativistic Astrophysics,
\item Bei-Lok Hu: Quantum Cosmology and Related Topics
\item Gary Horowitz: Interface with Mathematical High Energy Physics and
String Theory
\item Beverly Berger: News from NSF
\item Richard Matzner: Numerical Relativity
\item Abhay Ashtekar and Ted Newman: Mathematical Relativity
\item Bernie Schutz: News From Europe
\item Lee Smolin: Quantum Gravity
\item Cliff Will: Confrontation of Theory with Experiment
\item Peter Bender: Space Experiments
\item Jens Gundlach: Laboratory Experiments
\item Warren Johnson: Resonant Mass Gravitational Wave Detectors
\item David Shoemaker: LIGO Project
\item Peter Saulson and Jorge Pullin: former editors, correspondents at large.
\end{itemize}
\section*{Topical Group in Gravitation (GGR) Authorities}
Chair: \'{E}anna Flanagan; Chair-Elect: 
Dieter Brill; Vice-Chair: David Garfinkle. 
Secretary-Treasurer: Vern Sandberg; Past Chair: Jorge Pullin;
Delegates:
Bei-Lok Hu, Sean Carroll,
Vicky Kalogera, Steve Penn,
Alessandra Buonanno, Bob Wagoner.
\parskip=10pt

\vfill
\eject

\section*{\centerline
{Singularity avoidance in canonical quantum gravity}}
\addtocontents{toc}{\protect\medskip}
\addtocontents{toc}{\bf Research Briefs:}
\addcontentsline{toc}{subsubsection}{
Singularity Avoidance in Canonical Quantum Gravity, by Viqar Husain}
\parskip=3pt
\begin{center}
Viqar Husain,University of New Brunswick
\htmladdnormallink{vhusain-at-unb.ca}
{mailto:vhusain@unb.ca}
\end{center}

{\bf Singularity Avoidance in Canonical Quantum Gravity} \bigskip

The question of whether quantum gravity has something to say about spacetime curvature singularities has a long history, from the early investigations in the 1960's, to more quantitative work on symmetry reduced models in the 1970's using the Wheeler-DeWitt quantization. Some examples of this include work by Misner [1], and by Blythe and Isham [2] on Friedmann-Robertson-Walker models with a scalar field. 

After a lull of several years, the canonical quantum gravity programme was revitalised by the Ashtekar's triad-connection phase space variables for general relativity. The variables naturally represented as operators in this so-called loop quantum gravity approach is the densitized inverse triad, and the holonomy of its conjugate connection variable.   

The first significant development concerning singularity avoidance in the 
loop quantum gravity (LQG) programme occurred inadvertantly, in an attempt 
to define a regularized Hamiltonian constraint operator.  This was the 
construction of a  triad operator by Thiemann [3]. It was realized a few 
years later by Bojowald [4] that the algebraic relation used to define 
this operator in the full theory could also be used in quantum cosmology 
to define an ``inverse scale factor'' operator. The ``source'' of singularity 
avoidance in the symmetry reduced models studied so far in ``loop quantum 
cosmology'' (LQC) is that the inverse scale factor operator is bounded, at 
least in the isotropic models. 

The basic mechanism used in defining the inverse scale factor operator in LQC may be illustrated in a mechanical system. Consider a 2-dimensional phase space $(x,p)$ with planar topology, and consider a quantization on a spatial equi-spaced lattice with spacing $a$. There is a quantization and a basis such that the translation $e^{ia p}$ is realized as a shift operator, and the position $x$ is diagonal. An inverse $x$ operator can now be defined by starting with the Poisson bracket identity 
$$
       {1\over|x|} = -{4\over a^2} \left( e^{-iap} \left\{e^{iap},\sqrt{|x|}\right\}\right)^2,    
$$   
and writing an operator expression for the righthand side. Using the expressions for the translation and position operators, it is evident that this inverse position operator is diagonal in the position basis, and is bounded. This is the essence of the singularity avoidance result in LQC. 

The Poisson bracket ``trick'' used for the singularity avoidance result is applicable in other models for quantum gravity outside the LQG context; it does not depend on the use of the connection-traid variables which are at the basis of LQG, but rather on the choice of representation used for quantization. This observation was exploited by Husain and Winkler to revisit the quantization of models systems formulated using the ADM variables. This has led to singularity avoidance results for FRW models which are qualitatively similar to those found in LQC, and also for the gravitational collapse problem  in spherical symmetry [5]. 
 
There are two aspects of the results of singularity avoidance. The first is kinematical in the sense that the operators corresponding to the inverse triad in the model systems is bounded. The second is dynamical in that Hamiltonian evolution is well defined  and unitary beyond the point of the singularity -- that is evolution does not terminate there. This  feature of evolution in LQG based models arises due to its innate lattice structure: the Hamiltonian constraint acts discretely so that the Hamiltonian constraint condition is a difference equation rather than a differential equation as in Wheeler-DeWitt quantization. For regions away from the classical singularity, the difference equation behaves merely like a discretisation of the corresponding Wheeler-DeWitt equation, but for regions close to the singularity this is not the case, due partly to the incorporation of the inverse scale factor operator in the difference equation. 

 This second aspect was demonstrated explicitly by Ashtekar, Singh and 
 Pawloski in an FRW model coupled to a massless scalar field [6]. By 
 considering evolution using the scalar field as a time variable, they 
 showed using a numerical computation that a wave packet loses coherence 
 as it evolves toward the classically singular region, and after a bounce 
begins to regain coherence as it evolves away from this region.  

 Beyond model systems, there has been a detailed investigation by 
Brunnemann and Thiemann of triad operators in full LQG [7]. 
One central result here is that such operators are not bounded above in 
the 
full theory in a strict sense -- there are certain classes of states in the 
kinematical Hilbert space of LQG that lead to the result. While the 
inverse scale factor in anisotropic LQC models is also unbounded, its 
eigenvalues on zero volume eigenstates is bounded. In full LQG, however, 
even on zero volume eigenstates the operator corresponding to the inverse 
scale factor is unbounded.

 This work brings to the fore in the LQG setting the old question of the 
 relevance of mini-superspace quantization for understanding quantum 
 gravity. Specifically to the issue of singularity avoidance, it raises 
 the question of what features of the full theory are ultimately 
 responsible for this, and how it manifests itself in symmetry reduced 
models. For instance, the second conclusion drawn from [7] is that the 
expectation value of the inverse scale factor operator remains bounded in 
the sense of expectation values with respect to a one parameter 
family of coherent states whose peak in phase space follows the 
classically singular trajectory. This result of full LQG implies a 
completely different sense of singularity avoidance than the one obtained 
in LQC, but at least it does not contradict the LQC result.

 Further understanding of the issues raised by this work entail going 
 beyond the simplest models, to at least those that have some 
 inhomogeneity. This is being studied by several people both in the 
context of cosmology and gravitational collapse. 

\vskip 0.5cm

\begin{itemize}

\item{[1]} C. Misner, Phys. Rev. 186, 1319 (1969).  

\item{[2]} W. F. Blyth, C. J. Isham, Phys. Rev. D11, 768 (1975).

\item{[3]} T. Thiemann, Class. Quant. Grav. 15, 839 (1998) \htmladdnormallink{gr-qc/9606089}{http://arXiv.org/abs/gr-qc/9606089}

\item{[4]} M. Bojowald, Phys. Rev. Lett. 86, 5227 (2001) \htmladdnormallink{gr-qc/0102069}{http://arXiv.org/abs/gr-qc/0102069}
; Living Rev. Rel. 8, 11 (2005) \htmladdnormallink{gr-qc/0601085}{http://arXiv.org/abs/gr-qc/0601085}
.

\item{[5]} V. Husain, O. Winkler, Phys. Rev. D69, 084016 (2004)\htmladdnormallink{gr-qc/0312094}{http://arXiv.org/abs/gr-qc/0312094}
; Class. Quant. Grav. 22, L127 (2005) \htmladdnormallink{gr-qc/0410125}{http://arXiv.org/abs/gr-qc/0410125}
.

\item{[6]} A. Ashtekar, T. Pawlowski, P. Singh, Phys. Rev. Lett. 96, 141301 (2006) \htmladdnormallink{gr-qc/0602086}{http://arXiv.org/abs/gr-qc/0602086}
; Phys. Rev. D73, 124038 (2006) \htmladdnormallink{gr-qc/0604013}{http://arXiv.org/abs/gr-qc/0604013}
. 

\item{[7]} J. Brunnemann, T. Thiemann, Class. Quant. Grav. 23, 1395 (2006)\htmladdnormallink{gr-qc/0505032}{http://arXiv.org/abs/gr-qc/0505032}
; Class. Quant. Grav. 23, 1429 (2006) \htmladdnormallink{gr-qc/0505033}{http://arXiv.org/abs/gr-qc/0505033}
.

\end{itemize}

\section*{\centerline
{What's new in LIGO}}
\addtocontents{toc}{\protect\medskip}
\addcontentsline{toc}{subsubsection}{
What's new in LIGO, by David Shoemaker}
\parskip=3pt
\begin{center}
David Shoemaker,MIT 
\htmladdnormallink{dhs-at-ligo.mit.edu}
{mailto:dhs@ligo.mit.edu}
\end{center}

{\bf What's new in LIGO} \bigskip

Here is a brief update on the advances in LIGO -- our name for the
LIGO Laboratory (Caltech/MIT) and the greater LIGO Scientific
Collaboration (LSC).

\medskip
{\bf Observing}  The LIGO S5 science run, including the 4km- and
2km-length interferometers at Hanford, Washington, the 4km
instrument at Livingston, Louisiana, and the GEO600 Detector near
Hannover, Germany, continues. There have been some brief breaks for
minor commissioning and repairs, and happily both the sensitivity
and the uptime of all the instruments have been improving through
the run. The LIGO instruments are now exceeding their sensitivity
requirement by about a factor of three, and effectively meeting the
goal sensitivity curve laid out in 1995. Our commitment to the NSF
is to collect one integrated year of data with the instruments at
their design sensitivity, and we have now roughly 40\% of those
data. The German/UK GEO600 detector is also working nicely, with a
very high duty cycle.

\medskip
{\bf Analysis} The analysis pipelines are continuing to be refined,
and we are catching up on the continuous data stream from the
instruments. Results are in preparation from the S4, and the first
half of the S5 run. Recently published papers on data analysis are
``Search for Gravitational Wave Bursts in LIGO's Third Science Run''
(B. Abbott et al. (LSC), Class. Quantum Grav. 23, S29-S39 (2006);
\htmladdnormallink{gr-qc/051146}{http://arXiv.org/abs/gr-qc/0511146}
) and ``Upper Limits on a Stochastic
Background of Gravitational Waves'' (B. Abbott et al. (LSC), Phys.
Rev. Lett. 95, 221101 (2005); \htmladdnormallink{gr-qc/0507254}{http://arXiv.org/abs/gr-qc/0507254}
). Soon to
appear will be ``Search for gravitational wave bursts in LIGO's
third science run'' (\htmladdnormallink{gr-qc/0511146}{http://arXiv.org/abs/gr-qc/0511146}
).

\medskip
 {\bf Enhanced LIGO} Once the S5 run is complete, probably in
the Fall of 2007, the Collaboration will make some incremental
improvements to the detectors, principally to the interferometer
readout system. An increase of sensitivity of roughly a factor of 2
over a broad range of frequencies is anticipated, increasing the
volume of space searched by a factor of 8. We will run the
instruments once again for an extended run with this improved
sensitivity. The design for these improvements has made significant
progress over the last six months.

\medskip
 {\bf Advanced LIGO} Our plans for significant further
improvements to the instrument sensitivity were reviewed by the NSF
at the end of May. This `Baseline Review' is intended to complement
an earlier technical review, and covered the organization, costing,
schedule considerations, and readiness for project funding. The
review committee stated in its written review that it was
``impressed'' with the plans for the upgrade, and we feel we have
reason to be hopeful that the NSF, OMB, and Congress will find that
this is a good use of the taxpayers' money. The Advanced LIGO
instrument will have more than a factor of 10 better sensitivity
than the Initial LIGO instruments now running, increasing the number
of candidate sources by more than 1000, and should make observation
of gravitational-wave sources a common event. We plan to start the
project in 2008, start installing the new instruments in 2010, and
be collected interesting data in 2014.

\vfill\eject

\section*{\centerline
{Scanning New Horizons: GR Beyond 4 dimensions}}
\addtocontents{toc}{\protect\medskip}
\addtocontents{toc}{\bf Conference reports:}
\addcontentsline{toc}{subsubsection}{
Scanning New Horizons: GR Beyond 4 dimensions, by Donald Marolf}
\parskip=3pt
\begin{center}
Donald Marolf, UC Santa Barbara 
\htmladdnormallink{marlof-at-physics.ucsb.edu}
{mailto:marolf@physics.ucsb.edu}
\end{center}

For ten weeks this winter, a diverse collection of gravitational
physicists gathered at Santa Barbara's Kavli Institute of
Theoretical Physics to discuss myriad phenomena related to gravity
in $d \neq 4$ spacetime dimensions.  Though there was also
interest in the case $d=3$, the program (organized by Luis Lehner,
Rob Myers, and myself) largely focused on the case $d> 4$.   Why
study gravity outside of four dimensions?  In my opinion, the most
basic reason is that the dimension is a parameter one can dial to
learn more about the deep nature of gravitational phenomena.
Recall, for example, that via Kaluza-Klein reduction a theory with
$d > 4$ can in some cases be regarded as a 4-dimensional theory
with complicated matter fields.  Thus, one might expect any truly
universal property of gravity to apply equally well to both high
and low dimensions.  The Bekenstein-Hawking area law for black
hole entropy is perhaps a prime example such such a
dimension-independent phenomenon, and as such is widely regarded
as a deep principle of gravitational physics.  One aim of this
program was to discover what other phenomena are similarly
universal, and which phenomena are not.  Other motivations for
studying higher dimensional gravity include string theory, various
`large extra dimension' scenarios for our universe, and general
fun with mathematical physics.

Gravity in higher dimensions exhibits a number of striking features which stretch our 3+1 intuition.  For this reason, a primary goal of the program was to bring together higher-dimensional physicists with specialists (e.g. mathematical physicists or numerical physicists) who usually work in 3+1 dimensions.  The hope was that by bringing to bear sophisticated tools, progress could readily be made on a number of higher dimensional issues, mostly centered on the physics of black holes.  As usual in general relativity, these issues focused on existence, uniqueness,  thermodynamics, stability, and dynamics.

This fusion was quite successful, and the spectrum of results which came out of the workshop is too broad to fully summarize here.  With apologies to those whose work I will not mention (and to those for whom I mention only a small part of their work), I'd like to quickly review a few areas which were the focus of much discussion and where progress was especially significant, and/or there is great potential for further input from inspired readers of this article.  I emphasize that only a small fraction of the interesting results obtained are mentioned below.

\medskip

{\bf Existence, uniqueness, and thermodynamics}

What sorts of black objects exist in higher dimensions?  Recently, we have learned that higher dimensions host a rich spectrum of black objects, including stationary black rings for which cross-sections of the horizon are not spheres; see [1] for a recent review.    In 3+1 dimensions, Hawking's theorem guarantees that the horizon topology is spherical.  Using similar methods, some results on the higher dimensional case were already known.  However, a forthcoming set of papers ([2] and others to appear) by Lars Andersson, Greg Galloway, Jan Metzger, and Rick Schoen  will close an important loophole related to the possibility of Ricci-flat metrics on the horizon,  further narrowing the possibilities.  

As with Hawking's theorem, it can be quite unclear how a given result will generalize to higher dimensions.  A particularly interesting example is that of the black hole rigidity theorem, which states that every stationary black hole is axisymmetric; i.e., that it has a rotational Killing field.  An important corollary of this result is that the event horizon of a stationary black hole is a Killing horizon, and thus that it has a well-defined surface gravity which is constant over the horizon.  In this way, the rigidity theorem is deeply connected to black hole thermodynamics, and one would expect it to generalize readily to all dimensions.  However, the standard proof in 3+1 dimensions [3] relies on the fact that cross sections of the horizon have topology $S^2$.  The generalization to higher dimensions is highly nontrivial, and has only recently been established by Hollands, Ishibashi and Wald [4], in part as a result of the KITP program and interaction with Jim Isenberg and Vince Moncrief, from whom a related paper is expected soon.

\medskip

{\bf Stability}

Stationary black rings exist, but are they dynamically stable?  A
number of potential instabilities have been discussed in the
literature: radial instabilities, Gregory-Laflamme Instabilities
(see below), super-radiant instabilities, and potential
instabilities associated with absorption or emission of radiation.
Our program saw significant progress in establishing that black
rings {\it do} suffer from such instabilities.  First, a work by
Jordan Hovdebo and Rob Myers established [5] that very large black
brings are unstable to a Gregory-Laflamme type instability.  In
addition,  Henriette Elvang, Roberto
Emparan and Amitabh Virmani provided evidence [6]  that all neutral black rings (at least, in 4+1
dimensions) are unstable in some way.  For the branch of the black
ring solutions with the largest entropy, they show that such black
rings suffer from a radial instability as well as a
Gregory-Laflamme instability.  Furthermore, they give evidence
that the Gregory-Laflamme instability should lead the black ring
to break up into a set of black holes with large {\it orbital}
angular momentum.  Finally, Oscar Dias showed [7] that if `doubly
spinning' black rings exist, then they will necessarily have a
super-radiant instability.

\medskip
{\bf Dynamics}

A new dynamical issue that arises in $d > 4$ dimensions is the
Gregory-Laflamme instability [8] of thin black strings, membranes,
etc.  It is known that many black string solutions are linearly
unstable to perturbations that break translational symmetry along
the string but which preserve rotational symmetry around the
string.  The instability causes the string to become `lumpy,'
thickening in some places while thinning at certain `necks.' The
endpoint of this instability has been a subject of much debate and
discussion.  The original work [8] conjectured that the thin necks
might shrink to zero size and then `break,' so that the endpoint
is a set of separated black holes.  Much of the interest in this
work revolves around the fact that such a bifurcation would
violate certain forms of Cosmic Censorship.  Because the
interesting question involves the non-linear regime, it is natural
to explore this question numerically.  Indeed, the 2003 numerical
simulation [9] was the focus of much discussion at the program. In
particular, when plotted against their asymptotic time coordinate,
the evolution of many quantities in the spacetime shows signs of
slowing significantly near the point where their code crashes. One
might take this as evidence in favor of a scenario (see e.g. [10])
in which the endpoint is simply a static lumpy black string. It
goes without saying that better numerical simulations are needed
(and the authors of [9] are making progress in this direction).
However, our discussions also produced the conclusion that more
physics could be obtained by analyzing the data of [9] in terms of
a more physical time coordinate; e.g., the retarded time along
past null infinity. Because this retarded time might differ
substantially from the coordinate time of [9], such a new analysis
might suggest very different results more in line with the
original bifurcation suggestion of [8].   Interested individuals
may wish to consult [10,11,12] for more detailed discussions of
possible endpoints.

\medskip
{\bf Summary}

The 2006 KITP program `Scanning new horizons: GR Beyond 4
dimensions' was a period of intense interaction and discussion
between a diverse array of physicists which led to a number of
exciting new results. Yet much remains to be done, and in
particular there is much room for input from both mathematical and
numerical relativists. Although the physics concerns large numbers
of dimensions, interesting special cases often have sufficient
symmetry to reduce problems to either 3 or 2+1 dimensions or less.
Examples of such problems include the stability of
`ultra-rotating' higher dimensional black holes, the existence of
stationary black rings in $d > 5 $ dimensions, and the existence
and stability of `braneworld' black holes.  In such cases,
numerical analyses can be especially useful. I will be only too
happy if this short summary encourages others to enter this
exciting field and to further develop existence, uniqueness,
thermodynamic, stability, and dynamic results in gravity beyond 4
dimensions.   Links to the program talks and discussions can be
found at http://online.itp.ucsb.edu/online/highdgr06/, and provide
a useful introduction to many such topics.

{\bf References}

[1]   R.~Emparan and H.~S.~Reall,
  ``Black Rings,''
  \htmladdnormallink{hep-th/0608012}{http://arXiv.org/abs/hep-th/0608012}
.

[2]   G.~J.~Galloway,
  ``Rigidity of outer horizons and the topology of black holes,''
\htmladdnormallink{gr-qc/0608118}{http://arXiv.org/abs/gr-qc/0608118}
  .

[3] Hawking, S.W.: Black holes in general relativity. Commun. Math. Phys. 25, 152-166 (1972)

[4]   S.~Hollands, A.~Ishibashi and R.~M.~Wald,
  ``A higher dimensional stationary rotating black hole must be axisymmetric,''
\htmladdnormallink{gr-qc/o605106}{http://arXiv.org/abs/gr-qc/0605106}
.

[5] J.~L.~Hovdebo and R.~C.~Myers,
  ``Black rings, boosted strings and Gregory-Laflamme,''
  Phys.\ Rev.\ D {\bf 73}, 084013 (2006)
\htmladdnormallink{hep-th/0601079}{http://arXiv.org/abs/hep-th/0601079}
.

[6]   H.~Elvang, R.~Emparan and A.~Virmani,
  ``Dynamics and stability of black rings,''
\htmladdnormallink{hep-th/0608076}{http://arXiv.org/abs/hep-th/0608076}
.

[7]  O.~J.~C.~Dias,
  ``Superradiant instability of large radius doubly spinning black rings,''
  Phys.\ Rev.\ D {\bf 73}, 124035 (2006)
\htmladdnormallink{hep-th/0602064}{http://arXiv.org/abs/hep-th/0602064}
.

[8] R.~Gregory and R.~Laflamme,
  Phys.\ Rev.\ Lett.\  {\bf 70}, 2837 (1993)
\htmladdnormallink{hep-th/9301052}{http://arXiv.org/abs/hep-th/9301052}
.

[9]
  M.~W.~Choptuik, L.~Lehner, I.~Olabarrieta, R.~Petryk, F.~Pretorius and H.~Villegas,
  ``Towards the final fate of an unstable black string,''
  Phys.\ Rev.\ D {\bf 68}, 044001 (2003)
\htmladdnormallink{gr-qc/o304085}{http://arXiv.org/abs/gr-qc/0304085}
.

[10]   G.~T.~Horowitz and K.~Maeda,
  ``Fate of the black string instability,''
  Phys.\ Rev.\ Lett.\  {\bf 87}, 131301 (2001)
\htmladdnormallink{hep-th/0105111}{http://arXiv.org/abs/hep-th/0105111}
.

[11]  D.~Garfinkle, L.~Lehner and F.~Pretorius,
  ``A numerical examination of an evolving black string horizon,''
  Phys.\ Rev.\ D {\bf 71}, 064009 (2005)
\htmladdnormallink{gr-qc/0412014}{http://arXiv.org/abs/gr-qc/0412014}
.

[12]   D.~Marolf,
  ``On the fate of black string instabilities: An observation,''
  Phys.\ Rev.\ D {\bf 71}, 127504 (2005)
\htmladdnormallink{hep-th/0504045}{http://arXiv.org/abs/hep-th/0504045}
.

\vfill\eject

\section*{\centerline
{Quantum Gravity in the Americas III}}
\addtocontents{toc}{\protect\medskip}
\addcontentsline{toc}{subsubsection}{
Quantum Gravity in the Americas III, by Jorge Pullin}
\parskip=3pt
\begin{center}
Jorge Pullin, Louisiana State University
\htmladdnormallink{pullin-at-lsu.edu}
{mailto:pullin@lsu.edu}
\end{center}

The third edition of the ``Quantum gravity in the Americas'' workshop
was held at PennState on August 23-26. The previous instances of the 
workshop
were at Mexico City and the Perimeter Institute. The workshop was structured
into discussion sessions. Each session centered around a topic with one
or two ``rapporteur talks'' of 20-40 minutes each, and shorter talks
contributed by participants. About 50 researchers from the US, Mexico,
Canada and Uruguay attended the meeting.

The first day had an interesting session on new results on
the renormalization group applied to Einstein's theory, chaired by Max
Niedermeier and with a presentation by Frak Saueressig. New
results suggest that a fixed point may exist for Euclidean
Einstein gravity in perturbation theory and that the theory
somehow has dimensionality four at large scales and two at shorter
scales, a result that matches those of dynamical triangulations
and quantum geometry underlying loop quantum gravity.

The second session of the day was about loop quantum cosmology, chaired
by David Craig and with Parampreet Singh presenting. Newer formulations
of loop quantum cosmology not only include the earlier attractive results
of inflation and tunneling through the singularity but they also have
a clear prediction for when the quantum regime stops and the classical
one starts.

The first afternoon session was chaired by Ted Jacobson and centered on
cosmology and observations. Daniel Sudarsky spoke about how the quantum
fluctuations of the early universe can morph into matter perturbations
in cosmology and possible mechanisms for this to happen. Nelson Nunes
commented on further results of loop quantum cosmology with possible
observational implications.

The last session of the first day was on effective descriptions,
chaired by Martin Bojowald. Summarizing a very interesting
recent development, Laurent Freidel considered 3-dimensional
gravity coupled to scalar field and presented the scalar field
theory on a non-commutative space that results by integrating the
gravitational degrees of freedom. Radu Roiban discussed effective
field theories in string theory and discussed their limitations in
the regimes in which they predict occurrence of various types of
singularities.

On day two the morning started with a session on discrete approaches, 
chaired by
Rodolfo Gambini with talks by Luca Bombelli and myself. Luca gave an
overview of causal sets and I presented the new results we have on
uniform discretizations for quantum gravity

I chaired the next session on the physical sector of loop quantum
gravity.  Very nice talks by Jerzy Lewandowski on the status of the
Hamiltonian constraint and by Bianca Dittrich on the ``master
constraint program'' summarized our current understanding of the
dynamics of the theory.

The afternoon sessions started with Laurent Freidel chairing a session
on spin foams where Robert Oeckl and Florian Conrady spoke on issues
related to the path integral formulation of loop quantum gravity.

The last session of the day was on quantum geometry and matter,
chaired by Seth Major and with talks by Kevin Vandersloot and Mikhail
Kagan on effective descriptions in cosmology and inclusion of
inhomogeneities and by Fotini Markopoulou on a recent suggestion
of a possible link between the mathematics of braiding in
3-dimensions and the physics of fundamental particles of the standard
model.

Saturday started with a session on quantum gravity phenomenology,
chaired by Daniel Sudarsky and with a comprehensive review by
David Mattingly. The second session was chaired by Jerzy
Lewandowski on mathematical issues of loop quantum gravity and
talks by Jose Antonio Zapata and Daniel Cartin.

The afternoon session on Saturday was about the interface of
gravity with thermodynamics and cosmology, chaired by Warner
Miller and with talks by Ted Jacobson on the `derivation' of
Einstein's equations from non-equilibrium thermodynamics and by
Stephon Alexander on the possibility of the role of
gravitational waves in Baryogenesis.

The conference closed with a discussion and summary chaired by
Chris Beetle including a round table on black hole entropy and
a final summary by Abhay Ashtekar.

The small setting of the conference, combined with ample time
for discussions and the fact that the audience was technically
savvy on the field led to very nice discussions and clarifications
of points and left those attending with a crisp overview of this
rapidly developing field.

\vfill\eject

\section*{\centerline
{New Frontiers in Numerical Relativity, 2006}}
\addtocontents{toc}{\protect\medskip}
\addcontentsline{toc}{subsubsection}{
New Frontiers in Numerical Relativity, by Luciano Rezzolla}
\parskip=3pt
\begin{center}
Luciano Rezzolla, Albert Einstein Institute 
\htmladdnormallink{rezzolla-at-aei.mpg.de}
{mailto:rezzolla@aei.mpg.de}
\end{center}

Traditionally, frontiers represent a treacherous terrain to venture
into, where hidden obstacles are present and uncharted territories lie
ahead. At the same time, frontiers are also a place where new
perspectives can be appreciated and have often been the cradle of new
and thriving developments. With this in mind, the numerical-relativity
group at the Albert Einstein Institute (AEI) organised a workshop with
the goal of exploring and understanding these “New Frontiers”. The
workshop took place from July 17-21, 2006 at the AEI campus in
Golm, Germany. The meeting was focussed on the numerous issues that
occur in numerical relativity, such as: formulations of the
Einstein equations, initial data, multiblock approaches, boundary and
gauge conditions, and of course relativistic fluids and plasmas.

Almost 20 years since the homonymous meeting held at Urbana-Champaign
(``Frontiers in Numerical Relativity'', 1988), this meeting saw the
enthusiastic participation of a great part of the community, with 127
participants present (in 1988 there were 55) and with a large majority
being represented by students and postdocs, a reassuring sign of good
health for the community. The program was organised so as to have
few talks with ample time dedicated to discussions, which were then
continued over breaks, meals and late evenings. In addition, a whole
session spanning the last afternoon was dedicated to an
``unconstrained'' discussion which covered some of the most
controversial issues that emerged during the conference. During this
discussion, led by E. Seidel, particular emphasis was placed on the
need for systematic comparisons between waveforms generated by
different codes, as well as on the connection to the data-analysis
community.

A good overview of the conference can be found on the webpage of the
conference \texttt{http://numrel.aei.mpg.de/nfnr}, which contains the
list of the participants, a copy of the program and downloadable
version of the talks. Because of this, in what follows I will simply
report the highlights of the different thematic sessions which
composed the program.

\begin{itemize}

\item \textbf{Formulation of the Einstein equations}
\smallskip

  This session saw talks covering issues that go from pointing out
  clues about ``why do codes crash'' (C. Bona), over to the
  generalized harmonic gauge conditions in use by the Caltech/Cornell
  group (L. Lindblom), to the well-posedness and equivalence of
  different formulations and their relations when ``live'' gauge
  conditions are used (J.M. Martin-Garcia), to conclude with a
  prescription on how to deal with constraint violations in first-order
  evolution systems. Particularly interesting was also the progress
  report on the ability to perform numerical simulations of the tensor
  wave equation with pseudospectral methods and which represents the
  first step towards the solution of a maximally-constrained formulation
  of the Einstein equations (J. Novak).

\item \textbf{Initial-Value problem}

  This session covered a classical topic in numerical relativity: the
  construction of initial data for binary black hole systems. The
  talks focussed on solutions found with a parallel multigrid solver
  for binary systems with non-trivial spin combination (S. Hawley), on
  how to improve the Bowen-York prescription the initial data with 
  spinning black holes (M. Hannam), or on how to use matched asymptotic
  expansions to obtain approximate but hopefully more realistic binary
  black hole initial data (W. Tichy). Particularly interesting were
  also the progress reports about the use of ingenious coordinate
  transformations to build quasi-equilibrium configurations of
  arbitrary binaries (M. Ansorg) or on how to take properly into
  account spin in the construction of initial data for binary black
  holes with spins and in circular orbits (H. Pfeiffer). Both
  approaches showed the impressive accuracy of pseudospectral methods
  for this type of problems.

\item \textbf{Evolution of vacuum spacetimes}

  A lot of excitement preceded this session and it was all well
  motivated. A number of impressive results were in fact presented,
  some of them simply beyond (a realistic) imagination only a couple
  of years ago. Some of the results on the ``moving-punctures''
  prescription, which have been recently published, were presented
  in great detail (C. Lousto, J. Baker) and led to a lively
  discussion.  Equally interesting were the talks of other groups
  reporting their ability to now perform multiple orbits simulations
  of binary black hole systems when treated using moving punctures and a
  conformal traceless formulation of the Einstein equations
  (P. Marronetti, B. Br\"ugmann, F. Hermann, D. Pollney). Of topical
  relevance to the community engaged in puncture evolutions, was the
  recent work which studied the stationary slicing of puncture
  spacetimes and the behavior of fields at the puncture (Br\"ugmann,
  Pollney). Also rather impressive were the results on binary inspiral
  and merger carried out within a harmonic formulation of the
  equations either as second-order systems with finite-difference
  techniques (F. Pretorius) or as a first-order system with
  pseudospectral methods (M. Scheel). While the latter approach still
  needs to find an effective management of the domains at the time of
  the merger, the quality of the results presented for the inspiral
  has provided additional evidence of the accuracy of spectral
  methods. In addition, a useful comparison between the harmonic and
  conformal-traceless formulations was also presented as a first
  application of a newly developed code (B. Szil\'agyi). Very
  interesting work is also being done in areas beyond the binary black
  hole problem, with simulations of general singularities and the
  apparent validity of the BKL conjecture (D. Garfinkle), or the
  formation of naked singularities in the collapse of an
  ultrarelativistic fluid (M. Snajdr), or on a new prescription to
  smooth-out a singularity and perform stable and accurate simulations
  (E. Schnetter).

\item \textbf{Evolution of non-vacuum spacetimes}

  The large number of abstracts submitted to this session is an
  important indication that numerical relativity is not interested
  only in evolutions of pure-black-hole spacetimes and that a wider
  bridge towards numerical relativistic-astrophysics can be built. The
  session saw talks over a wide range of topics, from the analysis of
  the dynamical barmode instability and which provided a conceptual
  framework to determine why and when the instability is suppressed
  (G. Manca), over to the use of a spectral-methods code to study the
  behaviour of rotating and magnetized stars in quasi-equilibrium
  (S. Bonazzola), and to the modelling of radio images of Sgr A* using
  accretion disk simulations from a General Relativistic
  Magnetohydrodynamics (GRMHD) code on a Kerr background (S. Noble).
  Focus of a lot of attention were also simulations of gravitational
  collapse with talks on either the collapse of stellar cores to
  proto-neutron stars or of dynamically unstable neutron stars to
  black holes. More specifically, results were presented of 3D
  simulations of realistic stellar cores employing a
  finite-temperature equation of state and an approximate treatment of
  deleptonization (C. Ott, H. Dimmelmeier) as well as of 2D
  simulations of magnetized stellar cores in the test-field
  approximation (T. Font). Also, results were presented of 3D
  simulations of uniformly rotating neutron stars in which a novel
  technique avoided the use of excision and has allowed calculation of 
  the first complete waveform of the process (L. Baiotti), as well as
  of 2D simulations in full GRMHD of differentially rotating neutron
  stars, whose dynamics could be of help in modelling the engines
  powering short gamma-ray bursts (B. Stephens). Two newly developed
  codes were also presented which solve the equations of GRMHD either
  on a fixed black hole background and developed to model jet
  formation (Y. Mizuno), or on an arbitrary background and developed
  to extend the applications of the Whisky code to scenarios in which
  magnetic fields play an important role (B. Giacomazzo). Last but not
  least, a critical assessment was made of present techniques to
  handle surfaces and interfaces in relativistic hydrodynamics, and
  which will need to be improved for a future description of multiple
  fluids (I. Hawke).

\item \textbf{Multiblock techniques and AMR}

  This session saw talks on an area of numerical relativity which
  has grown rapidly in recent years and is expected to be of
  increasing importance. In particular, details were given about the
  multidomain pseudospectral collocation methods used by the
  Caltech/Cornell group to evolve spacetimes with black holes
  (L. Kidder) as well on 6-patches schemes with either overlapping or
  simply touching patches. In the first case details were presented on
  the way in which the patch ghost zones are ``synchronized'' by
  interpolation, on the tensor basis used in each patch, and on the
  handling of non-tensor field variables (J. Thornburg). Similarly,
  the main ingredients for the touching patches approach, such as
  high-order summation by parts finite-differencing operators and
  compatible dissipation operators, the use of penalty methods for the
  inter-block boundaries and adaptive time stepping, were also
  discussed in detail (P. Diener). Finally, results were presented for
  a new class of analytic solutions to the linearized Einstein
  equations for the Bondi-Sachs metric to be used as a testbed in a
  code employing stereographic coordinates and six angular patches
  (N. Bishop) and for an AMR code with fourth-order discretization in
  space and time and exploting compactification in space. Examples
  were given on the use of this code for the study of non-interacting
  massive Klein-Gordon field together with Yang-Mills-Higgs systems
  (P. Csizmadia).

\item \textbf{Boundary Conditions and perturbative methods}

  Recent work on another of the classical areas of research in
  numerical relativity, that which is interested in the definition of
  mathematically consistent and numerically accurate boundary
  conditions, was presented in this session. More specifically, a
  presentation was made on how to use trapping horizons to provide
  simple and satisfactory inner boundary conditions in black-hole
  spacetimes making use of the excision technique (E. Gourgoulhon). In
  addition, outer boundary conditions were the focus of several talks
  which addressed issues such as the definition of well-posed
  radiation-controlling boundary conditions for the harmonic
  formulation of the Einstein equations (O. Rinne), or the use of
  absorbing boundary conditions through a geometric approach
  (O. Sarbach) and the application of absorbing boundary conditions in
  examples of the linearized form of the Einstein equations
  (L. Buchman). Finally, results were presented on how to use
  black-hole perturbation theory and effective-one-body ideas to
  determine the gravitational-wave signal for the inspiral and merger
  of binary black-hole systems in the extreme mass-ratio limit
  (A. Nagar).

\end{itemize}

  In recognition of his important work in the field, the conference
  hosted a public lecture by Jimmy York on ``Dynamical Principles of
  General Relativity'', held at the picturesque Schlosstheater im
  Neuen Palais, within the premises of the Sans Souci Park in
  Potsdam.

  Talks given at the conference will appear as regular refereed
  articles in a special issue of CQG to be published in 2007, with
  M. Campanelli and L. Rezzolla acting as editors.

\vfill\eject

\section*{\centerline
{Teaching General Relativity to Undergraduates}}
\addtocontents{toc}{\protect\medskip}
\addcontentsline{toc}{subsubsection}{
Teaching General Relativity to Undergraduates,
by Greg Comer}
\parskip=3pt
\begin{center}
Greg Comer, St. Louis University
\htmladdnormallink{comergl-at-slu.edu}
{mailto:comergl@slu.edu}
\end{center}

On July 20 and 21, 2006 the AAPT held a Topical Conference at Syracuse 
University on Teaching General Relativity to Undergraduates. \ Why? \ Because 
the time has arrived to incorporate special and general relativity fully into 
the general physics curriculum. \ If you need to be convinced, consider that 
the equivalence principle works so well it is almost obscene, GPS fails when 
GR is ignored, gravitational red-shift is a fact, the Hulse-Taylor Binary 
Pulsar is producing gravitational waves, gravitational lensing exists, the 
expansion of the universe is essential for cosmological nucleosynthesis 
(which produced the lighter elements H, He, Li, etc), supermassive black holes 
in galactic centers appear to be the norm rather than the exception, and the 
Laser Interferometer Gravitational-Wave Observatory is near its design 
operation. \ Let us not forget that Dirac's great prediction of antimatter 
came about after he merged the physics that is spacetime with quantum 
mechanics. \ Although not a reason for inclusion in physics courses, it is 
remarkable that Einstein's $E = m c^2$ is an icon of popular culture, and no 
doubt recognized by more people than Newton's $F = m a$. \ As always, knowing 
{\em why} relativity should be incorporated is one thing, knowing {\em how} is 
an entirely different matter. 

This conference was my first introduction to the increasingly hot pursuit of 
pedagogically sound models and curricula for teaching relativity at the 
popular, high school, and undergraduate levels. \ I learned that one does not 
have to be a relativity expert to participate fully. \ In fact, one of the 
goals is to develop a curriculum that does not require such expertise (for the 
simple reason that we cannot expect schools to have a relativist on staff). \ 
Another goal is to streamline delivery of the mathematics of relativity so as 
to deliver the ``goods'' that the students want to study and that educators 
want to teach: black holes, gravitational waves, cosmology, and so on.

Pearson Addison-Wesley and Cambridge University Press provided participants 
with desk copies of five of their important GR textbooks. \ The authors of 
four of these texts, Jim Hartle, Bernard Schutz, and Edwin Taylor, were in 
attendance, and we were able to ``pick their brains'' and get first-hand 
accounts of their texts. \ Broadly speaking, most of the books reflect two 
different approaches: math first or physics first. \ The notable exception is 
a new text by Schutz which is designed for a general audience; namely, 
students who are taking their first (and perhaps only) physics course. \ The 
math first approach develops the mathematical foundations (tensors and 
differential geometry), introduces the Einstein equations, and then provides 
applications. \ In the physics first approach, applications occur first and 
only after key physics concepts are encountered do the mathematical 
underpinnings and Einstein equations appear. \ Understanding of gravity as a 
curved spacetime phenomenon is acquired via analysis of specific solutions to 
the Einstein equations (such as those for black holes, gravitational waves, 
and cosmology). \ Loosely speaking, we can think of math first as going from 
the general to the specific, and physics first as the other way around. 

The speakers were well chosen, their presentations were fascinating, and the 
discussions afterward and in breakout sessions were captivating. \ My desire 
for teaching relativity was invigorated, and I came away truely optimistic 
about the possibilities. \ Jorge Pullin gave a very good review of the central 
ideas of relativity (without inundating us with complicated mathematics) for 
participants who were new to teaching a GR course. \ Jim Hartle presented his 
rationale for the physics first approach. \ He has found that students can 
understand many GR physical effects quickly if they have some knowledge of 
mechanics. \ He emphasized that the structure of a course on GR very much 
depends on the context of where it is delivered: Who will teach it? \ How much 
time is available? \ What is the target audience? \ Tom Moore, who has a 
remarkable wealth of experience in teaching relativity to undergraduates, 
discussed ways in which the mathematics of GR can be more easily grasped by 
the students. \ For example, he has found that basic concepts can be 
effectively presented via analysis of two-dimensional metrics, such as using 
non-Cartesian coordinates for the flat-space metric and exploring the metric 
properties of the sphere. \ His experience also shows that students need lots 
of drilling on index manipulation.  

While Neil Ashby and Rai Weiss talked seriously about GPS and observation and 
experiment in GR, respectively, they also gave us some really juicy tabloid 
tidbits. \ We learned from Neil Ashby that while GR corrections were available 
to GPS, the necessary circuitry was not turned on initially (maybe because a 
highly placed, powerful individual was not convinced the corrections were 
needed). \ After the satellites were in orbit, it was, more or less, 
immediately determined that the system was not working. \ Only after the GR 
circuits were switched on, did GPS live up to its promise. \ As they say in 
football: ``Score!'' \ Rai Weiss regaled us with his personal experiences of 
learning and teaching relativity and performing experiments that test GR. \ 
While preparing to teach relativity for the first time (in the Sixties), he 
spent some time studying the existing data. \ In his own inimitable style, 
he expressed his, shall we say, disappointment---OK, it was disgust---with 
what 
he found for, say, measurements of light deflection by the Sun. 

A major point that was emphasized several times is that there should be 
significantly expanded discussion of acceleration in special relativity. \ 
Even among professional physicists there is a common misconception (first 
created, apparently, by Einstein himself) that GR is needed to really 
understand acceleration. \ It was pointed out, however, that we have no 
conceptual difficulties with accelerations in elementary particle 
scattering, so why should we have a problem when applying it to, say, the twin 
paradox? \ Even better, Don Marolf showed how one can use acceleration in 
special relativity to understand salient features of horizons in GR. 

While Marolf's talk indirectly addressed the misconception about acceleration, 
Peter Saulson's talk was a direct rebuttal of another, which is the 
misconception that laser interferometry cannot be used to detect gravitational 
waves. \ The incorrect reasoning says there will be no interference because a 
passing gravitational wave will stretch and squeeze the legs of the 
interferometer and the light in the same way. \ Of course, analysis based on 
GR and the Maxwell theory is uneqivocal; there {\em is} interference. 

Finally, the talk of Stamatis Vokos had, perhaps, the most to say about roots 
of misconceptions of relativity. \ He and his colleagues have done rigorous 
studies on how students go about absorbing, processing, and applying what they 
learn in a relativity course. \ At the most basic level they have found that a 
student's understanding of simultaneity is crucial for learning relativity. 

The main message of the conference is clear: there is much work to be done, 
but Relativity will soon rise out of the abyss of undergraduate syllabus 
topics that are labeled ``Time Permitting.'' \ The call is out and a concerted 
effort is in place. \ If you want to help, now is the time to act. \ As a 
matter of fact, I have a personal request: I am the editor of a new section on 
relativity that is to be put together for the ComPADRE project. \ It is a 
web-based network of educational resources supporting teachers and students in 
physics and astronomy (see http://www.compadre.org/portal/index.cfm). \ Bruce 
Mason, a P.I.~of the project, visited the conference and spoke briefly  
about ComPADRE's basic philosophy, current status, and future goals. \ For 
those who would like to have their materials on relativity made available, 
please send an e-mail to comergl@slu.edu with (1) a link to the site, (2) 
the target students, or level of the course, and (3) a two or three sentence 
description. 
 
The AAPT was assisted by LIGO/Caltech, the NSF Physics Frontier Center for 
Gravitational Wave Physics at Penn State, and the Syracuse University 
Department of Physics. \ Finally, many thanks go out to the Organizing 
Committee for producing such an excellent meeting: Michelle Larson (Chair), 
James Hartle, Charles Holbrow, Dale Ingram, Richard Price, Peter Saulson, John 
Thacker, and Stamatis Vokos. \ To learn more about results from the workshop 
go to http://www.aapt-doorway.org/TGRU.htm. \ Workshop posters, slides from 
presenters' talks, and workshop proceedings can all be found there.

\vfill\eject

\section*{\centerline
{Ninth Capra Meeting on Radiation Reaction}}
\addtocontents{toc}{\protect\medskip}
\addcontentsline{toc}{subsubsection}{
Ninth Capra Meeting on Radiation Reaction, by Lior Burko}
\parskip=3pt
\begin{center}
Lior Burko, University of Alabama in Huntsville
\htmladdnormallink{burko-at-uah.edu}
{mailto:burko@uah.edu}
\end{center}

The Capra series of meetings (named after the ranch in Southern 
California---that Caltech alumnus director Frank Capra bequeathed to his 
alma mater---the venue of the first meeting in 1998) are annual meetings 
on radiation reaction, that focus on the finite-mass corrections to the 
motion of a small mass in the gravitational field of a much larger mass, 
and on the emitted gravitational waves. In addition to being an interesting 
fundamental problem in general relativity, it is also a timely one, as some 
of the most promising sources for low frequency gravitational waves, that 
can be observed by space borne detectors such as LISA, are the waves emitted 
when a stellar mass compact object inspirals into a supermassive black hole 
at a galaxy's center. With a typical mass ratio of $10^{-6}$--$10^{-5}$, the description of these sources and their waveforms are the main motivation for the Capra meetings. 
The Ninth Annual Capra Meeting was hosted by the Center for Gravitation and Cosmology of the University of Wisconsin--Milawaukee from June 9 to 11, 2006 (and followed by the now traditional ``post Capra Workshop" June 12 to 14), and organized by Warren Anderson, John Friedman, Eirini Messaritaki, and Alan Wiseman. Administrative support was provided by Steve Nelson. In addition to the host Center, the meeting was supported financially by the Graduate School at UWM and by the Center for Gravitational Wave Astronomy at the University of Texas at Brownsville. The slides presented at the meeting are available online at the meeting's website, {\tt http://www.lsc-group.phys.uwm.edu/capra9/}). The author of this Summary, and surely all the participants of this highly successful meeting, would like to thank the organizers, and especially Eirini Messaritaki, for all the hard work they have put in it. 

Seventeen talks were presented. A number of talks were about various aspects of radiation reaction for particle motion in the spacetime of a Kerr black hole: Carlos Sopuerta (work with Pablo Laguna at Penn State) discussed advances in numerical simulations of extreme mass ratio inspirals, using finite element methods to handle spatial derivatives, and finite differences for the temporal derivatives. While in Schwarzschild very good agreement (to $0.05 \% $) with finite difference methods is found \cite{sopuerta}, in Kerr grid density adaptability is still a problem, that leads to accuracy problems. 

One of the problems related to the so-called ``gauge problem" is that the full and singular parts of the gravitational field of a particle are often written in different gauges. One way to attack this problem is to find instead the Weyl scalars, which 
are gauge independent. Notably, the reconstruction of the metric perturbations from the Weyl scalars is problematic when a source particle is present, because there are more gauge conditions to satisfy than gauge degrees of freedom.  However, if one obtains the regularized Weyl scalars, then 
one has a solution of the {\em homogeneous} Einstein equation, and therefore one doesn't have the above problem in reconstructing the metric perturbations.  Then, the metric perturbations can be found in {\em any} gauge, because the Weyl scalars are gauge independent. 
Bernard Whiting (work with Larry Price at the University of Florida) reported on work in progress related to the finding of the metric perturbations in Kerr by first regularizing the Weyl  scalars, specifically the jump conditions on the radiative modes of the Weyl scalars. Whiting focused on circular but non-equatorial orbits in Kerr. Using this method Whiting found the leading regularization parameter ``$A$", and work is in progress on finding the other parameters. John Friedman (work with Tobias Keidl and Alan Wiseman at UWM) addressed a closely related question, of how to find the regularized Weyl scalars using a special gauge that is exploiting the separability of the Teukolsky equation. Keidl (work with Friedman, Swapnil Tripathi, and Wiseman at UWM) then applied this approach to a simple case of a static mass point in the Schwarzschild spacetime, and showed how to solve the Bardeen--Press equation for $\Psi_4$. For the case of a static electric charge in Schwarzschild, this method successfully reproduces the result of Smith and Will \cite{smith-will}. 

Dong--Hoon Kim (AEI) discussed work in progress on a mode sum calculation of regularization parameters in Kerr, specifically for scalar field self force for generic orbits in Kerr. Kim describes the singular field in THZ normal coordinates \cite{THZ}, and finds the regularization parameters ``$A$", ``$B$", and ``$C$". Katsuhiko Ganz (Kyoto University, work with Wataru Hikida, Hiroyuki Nakano, and Takahiro Tanaka) addressed the adiabatic evolution of orbits in Kerr with large inclination angles, extending previous work in \cite{ganz}.  Ryuichi Fujita (work with Hideyuki Tagoshi at Osaka University) discussed a new method to integrate the Teukolsky equation in the frequency domain, that is based on the MST formalism \cite{mano}, in which one expands the homogeneous solutions in hypergeometric functions near the horizon and in Coulomb wave functions near spatial infinity, and matches the solutions across an overlap region. Fujita applied the method to inclined orbits in Kerr with small eccentricity. For cases where fluxes are available from other methods, agreement to at least 6 significant figures is found. In Schwarzschild, agreement to 15 significant figures was reported \cite{fujita}. 

Other talks discussed issues related to a Schwarzschild black hole: Lior Burko (UAH) discussed the evolution of quasi-circular orbits under a local self force, including conservative effects \cite{burko}. 
Nakano (Osaka University, work with Norichika Sago, Hikida, and Misao Sasaki) 
discussed the solution of the metric perturbations in the Regge--Wheeler 
gauge, including a change to the asymptotically-flat gauge for the 
non-radiative modes, and reported on a post Newtonian expansion for the 
radial component of the self force for a particle in circular orbit. Hikida 
(Kyoto University, work with Sanjay Jhingan, Nakano, Sago, Sasaki, and 
Tanaka) discussed the change in the orbital parameters of a scalar charge's 
motion under a self force, and obtained the latter in an expansion in the eccentricity. Hikida found that for small values of the eccentricity, the conservative effects on the phases are small, and are not likely to accumulate to $\pi$. However, gravitational conservative self force effects are expected to be larger than the scalar field counterpart. Steve Detweiler (work with Ian Vega at the University of Florida) discussed the gravitational self force effects on geodesics in Schwarzschild, extending previous work on the scalar field counterpart \cite{detweiler}. Vega (work with Detweiler) described work in progress on second-order gravitational perturbations due to a point particle, that is based on replacing the point particle with a small black hole at the world line.

Abraham Harte (Penn State) reported on work based on Dixon's formalism to evaluate self forces on extended bodies, which was applied to electromagnetism in flat spacetime \cite{harte}. Eric Poisson (work with Roland Haas at the University of Guelph) described a method for calculation of regularized self forces based on a tetrad decomposition of the singular field \cite{poisson}. Instead of expanding a vector in vector harmonics, Poisson described its projection on an orthonormal tetrad, followed by an expansion in scalar harmonics. In a carefully chosen ``Cartesian" tetrad, only a finite number of coefficients are non-zero, which makes this formalism attractive. Haas described work in progress on the scalar self force for eccentric orbits in Schwarzschild based on a time domain calculation. Wiseman (work with Friedman, Keidl, and Tripathi at UWM and Samuel Gralla at Yale University and the University of Chicago) discussed the computation of the self force using a modification of  the Quinn--Wald axioms, using a method based on finding an exact mode sum that represents the entire singular field. Carlos Lousto (UTB) reviewed the recent exciting advances in numerical relativity \cite{lousto}.

\end{document}